\documentclass[conference]{IEEEtran}
\IEEEoverridecommandlockouts

\usepackage{algorithm}
\usepackage{algpseudocode}
\usepackage{amsmath,amssymb,amsfonts}
\usepackage{algcompatible}
\usepackage{setspace}

\usepackage{graphicx}
\usepackage{textcomp}
\usepackage{xcolor}
\def\BibTeX{{\rm B\kern-.05em{\sc i\kern-.025em b}\kern-.08em
    T\kern-.1667em\lower.7ex\hbox{E}\kern-.125emX}}

\begin{document}

\title{A Blockchain-Encryption-Based Approach to Protect Fog Federations from Rogue Nodes\\
}
\IEEEpubidadjcol
\author{\IEEEauthorblockN{Mohammed Alshehri, Brajendra Panda}
\IEEEauthorblockA{\textit{Department of Computer Science and Computer Engineering} \\
\textit{University of Arkansas}\\
Fayetteville, USA \\
{msalsheh,bpanda}@uark.edu}

}

\maketitle
%
%
\begin{abstract}
People have used cloud computing approach to store their data remotely. As auspicious as this approach is, it brings forth many challenges:  from data security to time latency issues with data computation as well as delivery to end users. Fog computing has emerged as an extension for cloud computing to bring data processing and storage close to end-users; however, it minimizes the time latency issue but still suffers from data security challenges. For instance, when a fog node providing services to end users is compromised, the users' data security can be violated.
Thus, this paper proposes a secure and fine-grained data access control scheme by integrating the Ciphertext Policy Attribute-Based Encryption (CP-ABE) algorithm and blockchain concept to prevent fog nodes from violating end users' data security in a situation where a compromised fog node is being ousted. We also classify the fog nodes into fog federations, based on their attributes such as services and locations, to minimize the time latency and communication overhead between fog nodes and cloud server. Further, the exploitation and integration of the blockchain concept and the CP-ABE algorithm enables fog nodes in the same fog federation to perform the authorization process in a distributed manner. In addition, to solve time latency and communication overhead problems, we equip every fog node with an off-chain database to store most frequently accessed data files for specific time, and with an on-chain access control policies table (On-chain Files Tracking Table) which must be protected from being tampered by malicious (rogue) fog nodes. Therefore, blockchain plays a vital role here as it is tamper-proof by nature. 
We demonstrate our scheme's efficiency and feasibility by designing algorithms and conducting a security analysis. The provided analysis shows that the proposed scheme is efficient and feasible in ousting malicious (rogue) fog nodes.

\end{abstract}

%
%
\section{Introduction}
Cloud computing is a thriving paradigm due to the enormous on-demand services it provides to end users over the internet. Cloud computing has provided customers with innovative features such as availability, scalability, and economy that help to satisfy the substantial demand for storage and computation resources \cite{fox2009above}.  End users outsource their data, to the core network on the cloud, for processing and storage.  However, there are many obstacles facing data owners. First, the response time between users and the cloud is high because the data is stored in far from the data owners. Second, end-users' data security and privacy are susceptible to violation because of the semi-trusted third party controls the cloud. The research community has studied the issues of data security and privacy in cloud computing by adopting and applying advanced cryptographic techniques. However, it is still demanding to invent a new technology to resolve the cloud latency issue \cite{ deng2015towards}\cite{ li2015ehopes}.

From here,  fog computing was introduced in 2012 \cite{ bonomi2012fog} to reduce the time latency between the cloud and end users' devices \cite{ deng2015towards}\cite{ li2015ehopes} and to provide new services and applications to end users. Fog computing paradigm is as small clouds close to end-user's devices, which allows customers to utilize the computing and storage services at the edge of the network \cite{ bonomi2012fog}. More specifically, fog computing comprises of multiple fog nodes which provide services to end users \cite{yi2015survey}. The fog computing paradigm has supported end devices with distinguishing features such as mobility, location awareness,  and low time latency \cite{stojmenovic2016overview}. Because fog computing is deemed as an extension of the cloud computing paradigm; therefore, it inherits some of the security and privacy obstacles in cloud computing\cite{takabi2010security}. In particular, a fog computing paradigm is susceptible to different threats, such as malicious attacks and technical issues. In this context, the end users' data traveling through fog computing nodes to the cloud is vulnerable to different violations. Therefore, it is necessary to design a scheme to protect end-users' data confidentiality, availability, and integrity by ousting the rogue (malicious) fog nodes. One reason for these vulnerabilities is that end users outsource sensitive data to the nearby fog node for processing and then to the cloud for further processing and storage. Because this fog node is seen as a connector between end users and the cloud, it is challenging to protect other fog nodes and the cloud from malicious attacks. End users' data security would be difficult to defend if the fog node providing services in terms of storage and computation is compromised or goes rogue. We have been motivated by those issues to protect customers' private data from being compromised. Also, a method to secure communication among fog nodes is required to exchange encrypted data for reducing the time latency to retrieve it from the far cloud.

In general, threats in the context of fog computing takes two forms. 1) data modification: if an adversary gets hands-on end users' private data, he/she might violate its confidentiality and integrity. Therefore, introducing a security scheme is necessary to prevent data confidentiality and integrity violation among fog nodes; and between fog nodes and the cloud. 2) unauthorized access: when an adversary compromises a fog node, he/she can get unauthorized access to the end users' private data. Besides, data availability is in danger of malicious violation in this way. Therefore, it is essential to introduce a security scheme to protect against rogue (malicious) fog nodes and to oust them off the network while maintaining low time latency feature fog computing provides. Additionally, it is vital to enabling the fog nodes to carry on a distributed authorization process and to communicate in a trust-less environment.

Attribute-Based Encryption (ABE) is a cryptographic primitive that enables one-to-many encryption. It comprises two types: ciphertext policy attribute-based encryption (CP-ABE) \cite{r1} and key policy attribute-based encryption (KP-ABE) \cite{ goyal2006attribute}.  CP-ABE was introduced by Bethencourt at el. \cite{r1} which is deemed as one of the most significant algorithms to provide fine-grained data access control.

Blockchain has attracted the attention of researchers since it was introduced in late 2008 \cite{nakamoto2008bitcoin}. Blockchain is a shared distributed ledger, which is secured, immutable, and transparent, that helps in processing and tracking resources without depending on a centralized trusted third party \cite{xu2017taxonomy}. With this promising technology, peer-to-peer nodes can communicate, and exchange resources where the decision is carried on in a distributed manner by the majority of the network's nodes rather than a single centralized trusted third party.

This paper proposes a scheme to oust rogue (malicious) fog nodes and to minimize the communication overhead between the cloud service provider (CSP) and fog nodes (FNs) by integrating the CP-ABE algorithm and blockchain technology. Blockchain is adopted as a medium to store on-chain tracking table to verify the identity of each fog node using the smart contract before they could access the encrypted data on the CSP.  The blockchain immutability feature prevents fog nodes from maliciously changing the on-chain tracking table, if such a change detected, the FN which issues the request is reported as a rogue fog node. 

The rest of this paper is organized as: Section (II) presents the related work. Section (III) focuses on the motivation of this paper. Section (IV) explains the proposed scheme, and the scheme description is broadly explained in section (V). Finally, the security analysis and the conclusion consecutively are in sections (VI) and (VII).  
%
%
\section{Related Work}
The main objectives of this paper are to propose a new scheme to allow fog nodes (FNs) to communicate in trustless environment, to maintain end users' data security, and to reduce the time latency and communication overhead between the cloud service provider (CSP) and the FNs by adopting and integrating the CP-ABE and blockchain. Therefore, we summarize the overview of any related state-of-the-art works in this section, a brief literature review about access control, rogue fog nodes, and fog nodes communication.
\subsection{Access Control}
Sahai et al. \cite{sahai2005fuzzy} proposed an attributes-based encryption (ABE) scheme that uses the identity-based encryption algorithm \cite{boneh2001identity}. Its two variants are the key-policy ABE (KP-ABE) and the ciphertext-policy ABE (CP-ABE) \cite{goyal2006attribute}\cite{r1}. ABE encryption and decryption processes are based on users' attributes, so the scheme was first adopted in cloud computing to help data owners overcome obstacles such as data security and data access control when outsourcing data to the cloud.

The concept of access control has been well researched in cloud-computing, but more investigations are still needed into fog computing. Recently, the research community has studied access control issues in this area and started to adopt ABE in the environment with the object of providing fine-grained data access control and guaranteeing data security.

ABE has been applied to IoT devices to resolve data access control issues. Yu et al. \cite{yi2015survey} presented the fine-grained data access control issues arising in a wireless server network, proposing an FDAC scheme in which each sensor node was assigned a set of attributes and each user was assigned an access structure to specify access rights.

Huang et al. \cite{huang2017secure} proposed a CP-ABE scheme to support data security and fine-grained data access control, using features such as ciphertext updating and computation outsourcing for fog computing with IoT devices. As Zuo et al. \cite{zuo2018cca} also found, their scheme's main problem is that it was only suitable with an AND-gate encryption predicate. Xiao et al.\cite{xiao2017hybrid} proposed a hybrid fine-grained search and access authorization scheme for fog computing, based on a CP-ABE cryptographic primitive. Their scheme was hybrid as it supported both index encryption and data encryption. This too, however, supported only AND-gate predicates. Mao et al. \cite{7087379} proposed the construction of a CPA-secure and RCCA-secure scheme; it was based on an ABE model with the possibility of outsourced encryption verification. Dsouza et al. \cite{salonikias2015access}proposed a scheme to support secure data sharing and communication in fog computing. This scheme is no more than a policy management framework as it lacks details on building the policy repository and  users' identities, as well as on making decisions and protecting users' identities and data privacy. Stojmenovic et al. \cite{ stojmenovic2016overview}studied authorization and authentication issues among fog devices and between fog and cloud. They based their study on the ABE algorithm and argued that end users could still be authenticated and authorized to fog devices even in the presence of a vulnerable connection between the fog and the cloud. Li et al.\cite{li2015robust} proposed a model to collect smart devices' attributes as dynamic attributes, incorporating them with the ABE algorithm to verify the access authority in real time. Mollah et al.\cite{ mollah2017secure} proposed a lightweight cryptographic model to provide access control and data-sharing; in this model, all security operations are offloaded to nearby fog servers.

\subsection{Rogue Fog Nodes}
A rogue fog node is a fog node that pretends to be legitimate and deceiving end devices, other fog nodes, or the cloud into communicating with it. Its malicious hidden intent is to violate owners' data security and privacy. Having compromised a fog node, an attacker can violate the security and privacy of end users' data in transit to the cloud through the fog nodes. The research community has not broadly addressed this problem. 

Stojmenovic et al. \cite{stojmenovic2014fog} proposed a scheme to show the feasibility of a man-in-the-middle attack in which the gateway is either being compromised or substituted by a fake. Han et al. \cite{han2009measurement}\cite{han2009measurement} proposed a measurement-based scheme to help users avoid connecting to fake access points. Their scheme discovers the rogue AP by calculating the round-trip time between end users and the DNS server. Mohammed et al.\cite{alshehri2019encryption} proposed an encryption-based approach to protect fog computing from rogue fog nodes based on the CP-ABE algorithm. In this scheme, fog nodes cannot exchange data unless they fully trust each other. So, to let the fog nodes to carry on a distributed authorization process is an outstanding idea.

\subsection{Communication among Fog Nodes}
Arwa et al. \cite{alrawais2017attribute} proposed an attribute-based encryption scheme to maintain fog communications security. Designed to provide authentic, confidential communications among a group of fog nodes, it made use of a key-exchange protocol based on the CP-ABE algorithm. The researchers have not widely addressed this issue. 

Recently, researchers have started using blockchain to develop secure applications, benefiting customers in various fields. Christidis et al. \cite{christidis2016blockchains} summarized the blockchain use cases built into the IoT. They reviewed the integration of a blockchain smart contract into the IoT. Kamanashis et al.\cite{ biswas2016securing} proposed a multi-layered security scheme to create a secure communication platform in smart cities by integrating blockchain with smart devices. Researchers use blockchain as a distributed database to store heterogeneous data related to the smart city, such as traffic and location. These data need to be shared among smart cities' components. The main issues dealt with by this scheme are scalability and reliability in smart cities. In\cite{7471347}, authors proposed a blockchain-based scheme, consisting of distributed data storage, to enable data sharing in IoT environments. The aim of using blockchain was to enable data access control and data storage. In their scheme, the authors separated data store from data management, then used blockchain to verify the separation and to decentralize the access control.

The fog computing layer occupies the middle ground between cloud servers and end users' devices and is, therefore, more susceptible to attack than other layers. Data owners can find their data damaged or leaked by a rogue fog node. Since one of the primary fog computing features is to increase network reliability, fog nodes need protection method against malicious attacks. Protecting the nodes will defend the data owners' security and privacy.

\subsection{Our Contributions}
To the best of our knowledge, no previous work has adopted and integrated the blockchain technology with the CP-ABE algorithm to form a fog federation (FF) and address the problem of rogue fog nodes in fog computing. Based on \cite{xu2017taxonomy} and \cite{r1}, we propose a secure scheme in the context of fog computing with the following features: 1) to detect and prevent rogue fog nodes from violating end users' data confidentiality and besides to minimize the communication overhead between the cloud server and the fog nodes. 2) to enable the fog nodes to communicate in a trust-less environment. 3) to enable the fog nodes to perform the authorization process in a distributed manner. 4) it supports end users' data availability.
However, we focus our work on the communication among fog nodes and between fog nodes and the cloud. We assume that the medium between the end users' devices and the fog nodes is secured for now.
\section{Motivation}
Fog computing plays critical roles in providing multiple services to end users such as music, adults' entertainment, kids' entertainment, emergency vehicle response, and health data collection at low latency time \cite{bonomi2012fog}. Even the innovative features fog computing paradigm provides, it introduces many security issues. However, customers cannot benefit from those services fog computing provides if the fog node, which provides the intended services, is not functioning thoughtfully. Therefore, costumers care about their data security and availability in different geographic location. This scheme is motivated by the idea of protecting customers' data security from rogue fog nodes. For example, in the health care sector, health care provider outsources patients' sensitive data to the close FN for processing and then for uploading to the cloud for permeant storage. In the future, suppose the FN responsible for these data is compromised, this FN could breach end users' data to attackers.
Moreover, end users would not be able to access his/her original data as the FN in charge of providing this service is comprised, so users need to be forwarded to another trusted fog node for data retrieving from the CSP. This paper focuses on preventing the security threats that rogue fog nodes cause the system to breach end users' encrypted data and the time latency to delivering the data from the cloud to the fog node.  Moreover, we leverage the blockchain technology to enable fog nodes to provide authorization in a distributed manner. So, we argue that this would help in recognizing rogue fog nodes. Specifically, an adversary can either take over the fog node or intercept data flown from the fog nodes to the cloud and vice versa. Based on the literature review, no previous scheme has addressed the issue of protecting the fog nodes from rogue nodes by combining a cryptographic primitives, such as CP-ABE, with blockchain. Hence, there is a need for an efficient and secure scheme that considers these security challenges and time latency.

%
%
\section{Proposed Scheme}

\begin{figure}[t]
\includegraphics[width=0.5\textwidth]{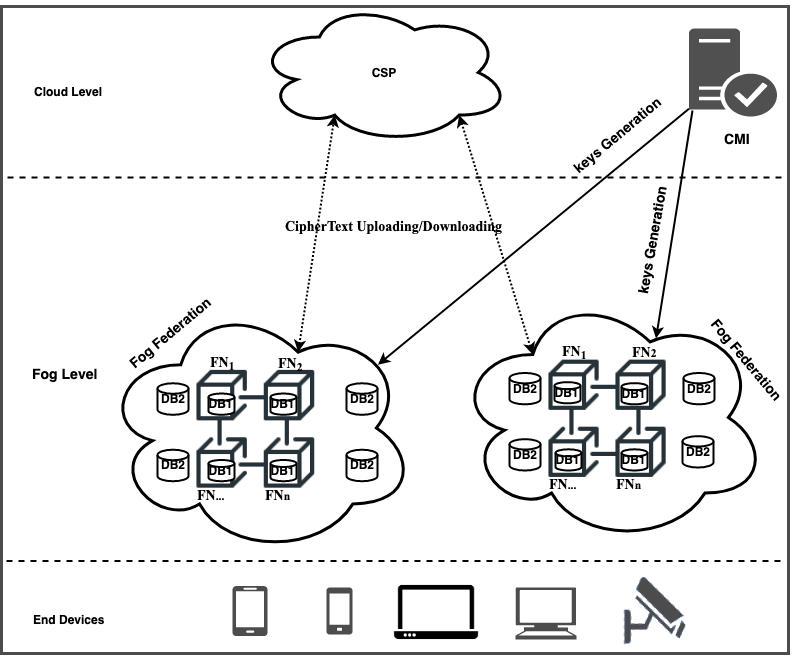}
\caption{The system model.}
\label{fig1}
\end{figure}

As Fig.~\ref{fig1} illustrates, our model comprises of the following entities: a cloud service provider (CSP), fog nodes (FNs), and a Cryptographic Materials Issuer(CMI). Fog nodes (FNs) with same attributes form a fog federation (FF) which is considered as a private blockchain. FNs in the FF function as miners to validate the requests generated by FN in same FF. Through this paper, we use fog federation and private blockchain interchangeably.

Data delivery from cloud storage to end users' devices is hindered by time latency and communication overhead. Fog computing has emerged to address these issues. However, FNs are exposed to malicious attacks; consequently, the end users' data security may be violated. To maintain confidentiality, integrity and availability for end users' data travelling through FNs to the CSP, we exploit and integrate blockchain with the CP-ABE algorithm, an advanced cryptographic primitive that enforces fine-grained data access control to secure communication between FNs and the CSP. Using blockchain, FNs can perform the authorization process in a distributed manner.

To solve the problem of high time latency and communication overhead, we equip every FN with an access control policies table (on-chain files tracking table),as shown in Table-\ref{tab4}, which must be protected from tampering conducted by malicious (rogue) FNs. Blockchain plays a vital role here, as it is tamper-proof by nature. 
This scheme equips each FN with a set of attributes, namely location and services. Each FN in the same FF maintains an on-chain files tracking table to provide fine-grained access control by using a smart contract. This allows other FNs in the same FF to access the encrypted data on the CSP when their attributes satisfy the stored predicates in the on-chain files' tracking table. When a FN sends a request to other FNs in the same FF to access a certain file, they check whether any FN in the FF has the required file stored in its off-chain database (DB). If so, they refer the requestor to get the file from the FN currently in possession of the file instead of sending the request to the CSP. This feature is a benefit of our scheme, as it minimizes the time latency and communication overhead between the CSP and the FF. In addition, the CSP maintains a verification list (VL) of FFs along with their Fog Federation Public Key $(FFPK)$ and Fog Federarion Attributes (FF$_{att}$), as shown in Table-\ref{tab1}.This table allows the CSP to verify the received request from FNs. It also stores each encrypted file (EF) along with a verification file (VF), as shown in Table-\ref{tab2}. The CSP uses this table to match each EF to the corresponding VF in order to verify the requests arriving from the FNs. 

\begin{table}[b]
\caption{Verification List}
\begin{center}
\begin{tabular}{|c|c|c|}
\hline
\textbf{FF$_{ID}$}&{\textbf{FFPK}}& \textbf{FF$_{att}$}\\
\hline
FF$_1$& Rtufn@10 &  (Health OR Education) and Atlanta \\
\hline
...& ... &  ... \\
\hline
FF$_n$& 1039gNF & Education and Atlanta   \\
\hline
\end{tabular}
\label{tab1}
\end{center}
\end{table}
\begin{table}[htbp]
\caption{Encrypted Files and Verification Files}
\begin{center}
\begin{tabular}{|c|c|}
\hline
\textbf{EF$_{ID}$}&{\textbf{VF$_{ID}$}}\\
\hline
F$_1$& VF$_1$ \\
\hline
...& ... \\
\hline
F$_n$& VF$_n$   \\
\hline
\end{tabular}
\label{tab2}
\end{center}
\end{table}
In fog computing, FNs collect data from end devices. They process those data and upload the results to the cloud for further processing and storage. We equip every FN with two databases (DBs), an off-chain and an on-chain DB. The off-chain DB stores the encrypted data most frequently accessed by end users, which reduces time latency when retrieving the encrypted data from the CSP. FN can retrieve data encrypted by FNs in the same FF. This feature helps maintaining data availability when ousting an FN from the FF near end users' devices. The on-chain database is considered as digital ledger and stored on the blockchain. It stores an on-chain file tracking table as shown in Table-\ref{tab4} to verify FF's members identifications. To protect cryptographic shares from breaching by malicious FN, it is stored off-chain by every FN, Table-\ref{tab3}. For the sake of clarity, we will consider the following example:

\begin{table}[b]
\caption{Cryptographic Shares}
\begin{center}
\begin{tabular}{|c|c|c|}
\hline
\textbf{EF$_{ID}$}&\textbf{VF$_{share}$}&\textbf{SK$_{share}$}\\
\hline
EF$_1$     &   MNGONMDF    &   SK$_1$ \\
\hline
...  &   ...  &   ... \\
\hline
EF$_n$     &   MNGONMDF    &   SK$_n$  \\
\hline
\end{tabular}
\label{tab3}
\end{center}
\end{table}
Suppose FN$_1$ encrypts file F$_1$ using a random secret key ($SK$). Then 
it generates a $VF$. Consequently, it uploads the EF into the CSP along with the VF and generates a new row in the on-chain file tracking table.

FN$_1$ divides the $SK$ and the $VF$ into n shares where n is the number of FNs in a given FF (private blockchain). It sends [$SK_{share}$, VF$_{share}$, EF$_{ID}$] to each FN in the FF to be stored as off-chain cryptographic shares, Table-\ref{tab3}.
All FNs in the same FF maintain an on-chain file tracking table, as shown in Table-\ref{tab4}. This table is used to verify the data requestor (FN$_{ID}$) by verifying the FN$_{sign-att}$. This table is protected from being tampered by malicious FNs through the nature of blockchain services. Suppose FN$_2$ in the FF seeks to retrieve an EF from the CSP. There are three scenarios, which are as follows.

In the first scenario, FN$_2$ needs to retrieve EF$_4$ from the CSP, and no FN has the file in its off-chain DB. FN(requestor) needs to send a request through the FF (private blockchain). Then, the FNs in the same FF would verify the requestor's attributes signature. If it is verified by the majority of the FNs in the FF, they would share the [$SK_{share}$, VF$_{share}$, EF$_{ID}$] they have related with the EF$_4$ with the requestor. When the requestor collects at least 51\% of the VF's shares, it sends a request to retrieve the EF$_4$ from the CSP as [VF$_{shares}$, EF$_4$,signed (FN{$_{att}$})]. Accordingly, the CSP verifies the VF$_{shares}$; if they match the VF attached with the EF$_4$, it sends the file to the FN to decrypt. For more details, read the scheme's description.

In the second scenario, FN$_2$ needs to retrieve EF$_2$ from the CSP, and FN$_1$ has the file in its off-chain DB. It sends a request through the private blockchain with the [EF$_2$, signed (FN$_{att}$)]. Accordingly, FNs verify the FN$_2$'s identity by verifying its signed attributes using $FFPK$. If verified, they check which of the off-chain DBs has the file and then send the FNs$_{ID}$ that have the EF$_2$ to the requestor. In this case, the file requestor, FN$_2$, does not have to contact the far CSP to retrieve the file. It could contact FN$_1$ to get EF$_2$. This feature gives our scheme a credit in minimizing the time latency and communication overhead between the CSP and the FFs. This is possible with the fog federation idea. For more details, read the scheme's description.

\begin{table}[t]
\caption{ On-Chain Tracking Table}
\begin{center}
\begin{tabular}{|c|c|c|c|c|c|}
\hline
\textbf{FN$_{ID}$}&{\textbf{EF$_{ID}$}}& \textbf{FN$_{sign-att}$}&\textbf{attributes set}&\textbf{off-chain DB}\\
\hline
FN$_1$ &   EF$_1$  &   fNJk3u@ & Movies and CL & EF$_2$ \\
\hline
FN$_2$ &   EF$_3$  &   KLJk3J@ & Edu and AT & 0  \\
\hline
... &   ...  &   ...  & ...  &... \\
\hline
FN$_n$ &   EF$_7$  &   @3EFJJL@   & Health and AR & 0  \\
\hline
\end{tabular}
\label{tab4}
\end{center}
\end{table}

In the third scenario, if FN (the requestor) is not verified, FNs in the FF will report the FN as a rogue FN and delete it from the on-chain tracking table. It then cannot access any encrypted data. 
 
An FN leaves the system for whatsoever reason, or it could go rogue after being compromised by a malicious attack. To ensure users' data security, namely data confidentiality, and availability, there has been a demand to take precautions against rogue FNs. Our scheme efficiently acts as a protection layer against rogue FNs to prevent them from accessing the encrypted data stored on the CSP.
 
All FNs in one FF maintain the same on-chain file tracking table. Malicious modification of the table by a malicious FN would be detected, as any update must be verified by the majority of the FNs, which is why we utilize the blockchain as FFs. If an FN issues an update and it is not passed by most of the FNs, the issuer would be considered as a rogue FN and then excluded from the FF. When a rogue FN is detected in the FF, it is reported to the CSP. Then, the CSP will not respond to any requests from this FN. Accordingly, this feature enhances the scheme's ability to protect against rogue FNs.
%
%
\section{Scheme Description}

\begin{figure}[b]
\includegraphics[width=0.3\textwidth]{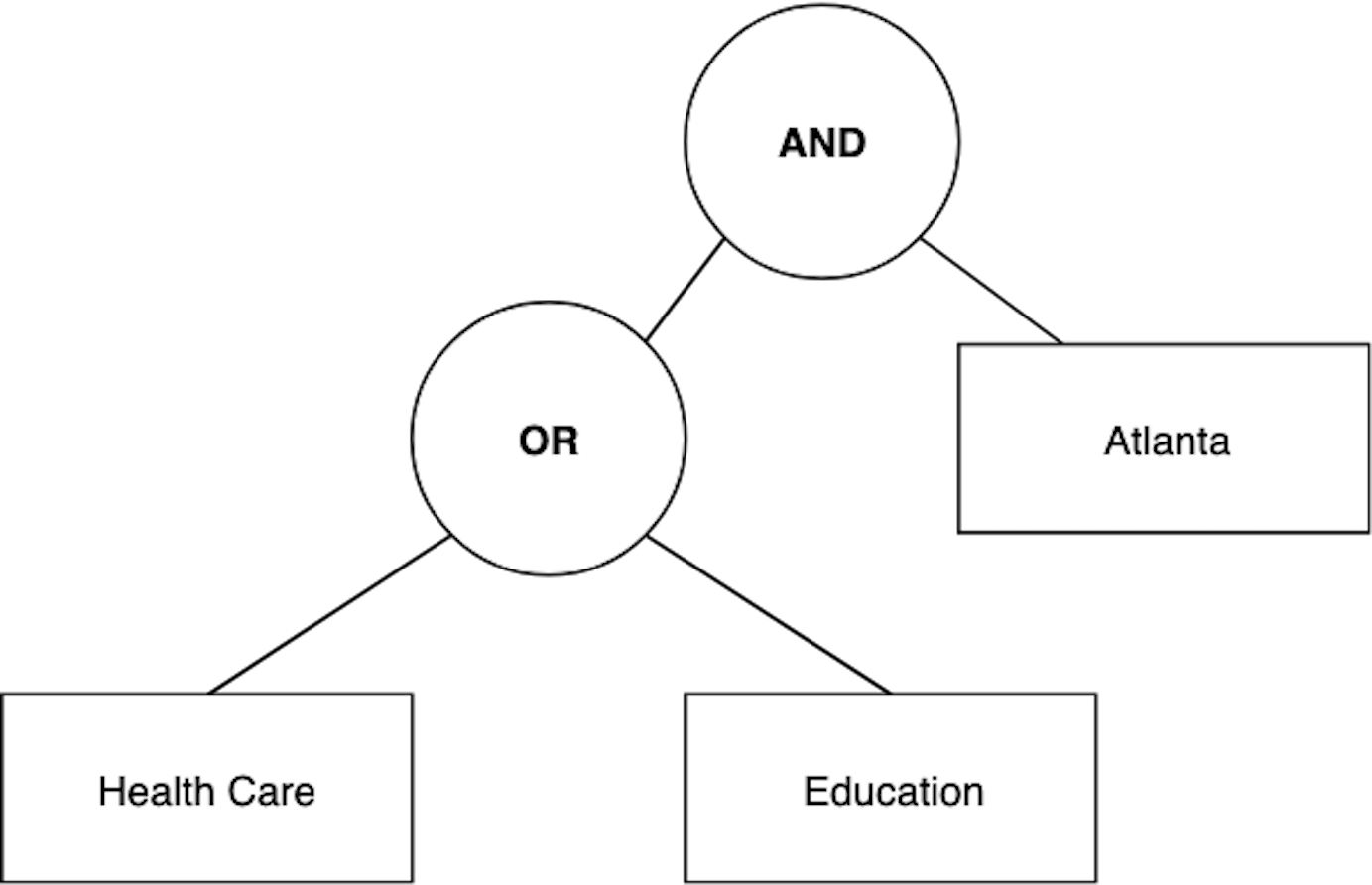}
\centering \caption{Example of an access structure.}
\label{fig2}
\end{figure}

In this section, we present the description of our scheme based on the integration of the CP-ABE algorithm and blockchain. We exploit the access tree model presented in \cite{r1} as an access structure A, as shown in Fig-\ref{fig2}. To satisfy the access tree conditions, we follow the same methods in \cite{r1}\cite{boneh2001identity}. For more information about the access tree model, read through \cite{r1}\cite{boneh2001identity}. Next we provide algorithms needed for our model.

\begin{algorithm}[t]
 \caption{ Setup($\lambda$)}
\begin{algorithmic}[1]
 \State Choose bilinear cyclic group $G_1$ of prime order $p$ with generator $g_1$;
 \State Randomly generates exponents $\alpha, \beta \in$ $G_p^*$ ; 
 \State Random Oracle $H$: $USAT \to$ \{0,1\}$^*$ // to map each att in USAT into random element $\in G_1$; 
 \State compute $SMK$ and $SPK$ as:
 \State\ \ \ \ \ \ \ \ \ \ \ \ \ \ \ \ \ \ \ \ \ $SMK$ = $\{\beta,g_1^{\alpha} \}$               
 \State\ \ \ \ \ \ \ \ \ \ \ \ \ \ \ \ \ \ \ \ \ $SPK$ = $\{G_1, g_1, g_1^\beta, e(g_1,g_1)^\alpha\}$ ;
 \State The $SPK$ is public to all FNs, but $SMK$ is private for CMI
\end{algorithmic}
\end{algorithm}

Algorithm 1, which is based on \cite{r1}, is executed by CMI. It receives a security parameter as an input, then it outputs a system master key ($SMK$) and a system public key ($SPK$). It builds a universal set of attributes (USAT), which FF classification is based on in our scheme; USAT = {FN's location, FN's services}.

\begin{algorithm}
 \caption{ Join (FN$_{att}$)}
\begin{algorithmic}[1]
 \State New FN contacts CMI to join the system
 \State CMI verifies the FN (requestor)
 \IF{not verified}
 \State Declined
 \ELSE \State Check the FFL
        \IF{FF$_{att}$ == new-FN$_{att}$}
        \State Assign new-FN$_{att}$ to the corresponding FF$_{ID}$
        \ENDIF
\State Assign new FN$_{ID}$ as: FF$_{ID}$.FN$_{ID}$ 
\State Send new FN$_{ID}$ to all FN in the same FF
\ENDIF
\FOR{every FN $\in$ FF}
    \State share on-chain ledgers (on-chain tracking table) with the new FN
\ENDFOR
\IF{New FN receives at most 49\% of un-identical on-chain identical ledgers}
\State report a problem to CMI
\ELSE
\State Return (on-chain ledgers are authenticated)
\ENDIF
\end{algorithmic}
\end{algorithm}

Algorithm-2 is executed by an FN that intends to join the system to provide a service for end users. It contacts the CMI, then the CMI checks the FFL. Then, it assigns the FN to the corresponding FF if its attributes are verified. If the FN is verified but no FF$_{att}$ matches its attributes, the CMI creates a new FF and adds the new FN to it. If the FN is not verified, the connection will be refused. However, as each FF is considered a private blockchain, the FNs already part of the FF have to share their on-chain ledger with the new FN. If the new FN receives 51\% of identical ledgers, it saves them in the on-chain DB. Otherwise, it reports an issue to the CMI. The details of how to verify and authenticate the FN's attributes are out of our scope for now.

\begin{algorithm}[t]
\caption{Generating $FNSK$ ($SPK,SMK,S,FN_{ID}$)}
\begin{algorithmic}[1]
\State CMI gets request from new FN to join the system
\State Choose bilinear cyclic group $G_1$ of prime order $p$ with generator $g_1$;
 \State Randomly generates exponents $\alpha_1, \beta_1 \in$ $G_p^*$ ; 
 \State Random Oracle $H$: $USAT \to$ \{0,1\}$^*$ // to map each att in USAT into random element $\in G_1$; 
 \State compute $FFPK$ and $FFMK$ as:
 \State\ \ \ \ \ \ \ \ \ \ \ \ \ \ \ \ \ \ \ \ \ $FFPK$ = $\{\beta_1,g_1^{\alpha_1} \}$               
 \State\ \ \ \ \ \ \ \ \ \ \ \ \ \ \ \ \ \ \ \ \ $FFMK$ = $\{G_1, g_1, g_1^{\beta_1}, e(g_1,g_1)^{\alpha_1}\}$ 
\State generate random $r \in Z_p^*$
\FOR{every $att_i \in S$}
\State generate random $r_i \in Z_P^*$
\ENDFOR
\FOR{each new FN$_i \in FF$}
\State compute $FNSK$ 
\State send $FNSK$ to FN$_i$
\ENDFOR
\end{algorithmic}
\end{algorithm}

Algorithm 3 is executed by the CMI. It generates a Fog Federation Public Key ($FFPK$) and a Fog Federation Master Key ($FFMK$). The $FFPK$ is public to all FNs in the FF, whereas the $FFMK$ is private to the CMI. It also generates a distinct Fog Node Secret Key ($FNSK$) for each FN in the FF. Then, the CMI securely shares the $FFPK$ and $FNSK$ with the new FN.
\begin{algorithm}[t]
\caption{Data Encryption ($M$, T, $SPK$, $FFPK$) }
\begin{algorithmic}[1]
\State SK = gen.DEC-Key
\State verification-file = generate.VF
\State VF$_{share}$ = VF/n    // n is the number of FNs in FF
\State divide SK key to shares as: SK/n
\State Encrypted-File (EF) = Encrypt (M, SK) //M is the message to encrypt
\State A is set of atts represented by monotone access structure tree T
\FOR{each node x in T}
\State set a polynomial
\IF{node x is root node in T}
\State set $q_{(0)R}$ = s // s is random value $\in Z_P^*$
\ELSE 
\State $q(0)_x$ = $q_{parent(x)}(index(x))$
\State then choose $d_x$ = $k_x$ - 1 as polynomial degree to interpolate with $q_x$
\ENDIF
\ENDFOR

\FOR{each y $\in$ Y}   //Y is the set of leaf nodes in T
\State $C_{y} = g^{q_{y}(0)}$
\State $C_y^{'} = H(att(y))^{q_{y(0)}}$
\ENDFOR
\State Compute: $C=g^{\beta s}$
\FOR{each $SK_i \in SK$}
\State Compute $C_{{ver}_i} = SK_i.e(g,g)^{\alpha s}$
\ENDFOR
\State Upload [EF, $EF_{ID}$, VF, FN$_{att_sig}$] to CSP
\FOR{each FN $\in$ FF}
    \State send [EF$_{ID}$, VF$_i$, $C_{{ver}_i}$ ]
\ENDFOR
\end{algorithmic}
\end{algorithm}

\begin{algorithm}[b]
\caption{on-chain tracking table smart contract }
\begin{algorithmic}[1]
\State sig = sign(FN$_{att}$, $FNSK$)  // for future authentication
\State propagate the transaction through FF [EF$_{ID}$, Current-Hash, Prev.Hash, Timestamp, FN$_{att-sig}$]
\IF{transaction is verified and FN$_{att.sig.verify}$ is true}
    \State File$_{owner}$ (FN) generates a new row in on-chain file tracking table as:[FN$_{ID}$, EF$_{ID}$, attributes-set, off-chain DB]
    \State FNs $\in$ FF update the on-chain tracking table
\ELSE
    \State decline and Report FN as rogue FN to all FNs in FF by its ID
\ENDIF
\end{algorithmic}
\end{algorithm}

Algorithm 4 is executed by authorized FNs in a given FF. The FN intending to encrypt a file generates a secret key (SK) and uses it to encrypt the file. It also generates a verification file ($VF$) which is used for further requestors verification by the CSP.
Then, it divides the $SK$ and the VF into n shares as n is the number of FNs in the FF. After that, it encrypts the $SK_{i,shares}$ using the CP-ABE algorithm to enable every FN in the same FF to decrypt the encrypted$_{SK}$. Finally, the data owner (FN) uploads the EF along with the VF to the CSP. We assume that the number of FNs in an FF is fixed for now.

Algorithm 5 is an on-chain smart contract which is triggered among authorized FNs in the FF. The basic function of this smart contract is to keep tracking which FNs have which files in the off-chain DB. This means less time latency to retrieve EFs from the CSP and less communication overhead between the CSP and the FFs. No FN can retrieve the EF or perform any operation unless it is authorized by the majority of the FNs in the FF. This is attributed to the blockchain tamper-proof feature. When a FN submits a rquest to retrieve an EF from the cloud it has to sign its attributes using its own $FNSK$, then the others FF's members will authenticate it by checking its signature using the $FFPK$.
This feature is a credit to our scheme, as it prevents FNs in the FF from falsifying information about files' tracking table or accessing the EF in the CSP. 
\begin{algorithm}[t]
\caption{File Retrieving Smart Contract}
\begin{algorithmic}[1]
\State \textbf{ Part 1: } FN (requestor) ID verification
\State FN send a request through FF as [FN$_{ID}$, EF$_{ID}$, FN$_{att-sig}$]
\FOR{every FN$_i \in$ FF}
\IF{FN$_{att-sig}$ == true}
\State send [$SK_{share}$, VF$_{share}$, EF$_{ID}$]
\ELSE
\State decline
\ENDIF
\ENDFOR
\IF{FN (requestor) recieves 51\% of (SK,VF) shares}
\State send (SK,VF) shares to CSP
\ENDIF 
\State \textbf{ Part 2: } CSP checks the VF$_{shares}$
\IF{VF$_{shares}$ match at least 51\% of VF}
\State send EF to the FN (requestor)
\ELSE
\State decline the request and report FN to CMI as rogue node
\ENDIF
\end{algorithmic}
\end{algorithm}

Algorithm 6 is a smart contract which two parts. First part is executed between the FN that intends to retrieve an EF from the CSP and FF's members. Second part is between the FN (requestor) and the CSP. FN first propagates a request to retrieve an EF from the CSP through the private blockchain. All FNs in the FF verify the requestor attributes by verifying the requestor attribute signature. When majority FNs verify the requestor (see algorithm 7), they send [$SK_{share},VF_{share}$]to the EF requestor. When the EF's requestor (FN) receives 51\% of the shares, it send [$SK_{share}$, VF$_{share}$, EF$_{ID}$] to the CSP. The CSP checks the shares against the pre-stored VF. If the shares match at least 51\% of the VF, the CSP sends the EF to the FN. If the shares do not match the VF, the CSP will report the requestor as a rogue FN to the CMI. This feature strengthens our scheme in ousting rogue FNs to protect end users' data confidentiality, availability and to authenticate the FF's members identity. 
\begin{algorithm}[t]
\caption{Consensus approach}
\begin{algorithmic}[1]
\State assume 3f + 1 of FNs in FF  // f is max number of FNs which may fail
\IF{2f +1 of FNs confirm the transaction}
\State FNs come to consensus
\ELSE
\State discard the transaction
\ENDIF
\end{algorithmic}
\end{algorithm}

Algorithm 7 is based on Practical Byzantine fault tolerance approach (PBFT) \cite{castro2002practical}. We assume that the FF has 3f + 1 of FNs, where f is the maximum number of FNs that could fail. To confirm a specific transaction in our scheme, at least 2f + 1 of FNs must agree on it. In this method, the authorized FNs in the FF come to a consensus state.

%
%
\section{Security Analysis}

\subsection{Confidentiality}
The EFs in the CSP is protected against rogue FNs. To retrieve an EF, the FN must sign its attributes using its $FNSK$, then send it with the request as a transaction through the private blockchain.
The EF's requestor must accumulate at least 51\% of [$SK_{shares}$, VF$_{shares}$] to be able to retrieve the EF from the CSP. This will not happen unless the EF's requestor passed the authorization process (algorithm 6) through the private blockchain. If the EF's requestor is not authorized by the majority of FF's members, it is reported as a rogue FN and consequently cannot access the EFs in the CSP or in an off-chain DB of other FNs. This feature makes our scheme efficient in protecting end users' data confidentiality.

\subsection{Availability}
The proposed scheme guarantees the data availability: even the FN providing services to end-users is down. The idea of the FF enables data owner to retrieve their encrypted data through any FN $\in$ same FF. 
The most frequently accessed EFs are stored in the off-chain DB of the FN that recently accessed it. This means the end users' data would be available near them for a period of time. The off-chain DB is flushed regularly as precaution against malicious FN.
\subsection{Ousting Rogue FN}
This scheme aims to oust rogue FNs from the system to keep end users' data secured from being breached or compromised. This feature is obvious through algorithms (5,6). In particular, if the FN (requestor) does not sign its attributes with the correct $FNSK$, FN authorization process will be denied by the private blockchain. Therefore, it would be ousted and protected from accessing the EFs in the CSP and in an off-chain DB of other FNs in same FF.
%
%
\section{Conclusion}
If the FNs providing services to end users are malfunctioning, end-users' data security can be violated. Also, it is possible that the FNs to go rogue which maliciously would violate end-users' data security. Besides, the time latency and communication overhead between the CSP and FNs are high when every time FN retrieves encrypted data from the CSP for processing.
This paper proposed a novel scheme to protect end-users' data security from being breached by rogue FNs and to reduce the time latency and communication overhead between the FNs and the CSP. It also enables FNs to communicate in a trust-less environment, and to perform authorization processes in distributed manner to detect the rogue fog node.
The proposed scheme exploited and integrated the CP-ABE algorithm and the blockchain concept. We classified the FNs into FFs (private blockchain) according to their attributes and equipped each FN with on-chain and off-chain databases.
The data requestor (FN) must sing its attribute using its own $FNSK$ before propagating a request through the private blockchain to access EF stored in the CSP. In case the requestor FN$_{att-sig}$ is not verified by the majority of the FNs $\in$ FF, the FN would be reported as a rogue FN. We provided a security analysis to show the proposed scheme is secured against rogue FNs and is efficient in reducing time latency and communication overhead between the CSP and the FNs. Our future work will conduct a simulation stage to test our scheme performance.
%
%

%
%

%
%

%

\vspace{12pt}

\end{document}